\documentclass{aa}
\usepackage{graphics,color}
\begin{document}
   \title{Long period variables in the globular cluster 47 Tuc:
radial velocity variations}

   \author{T. Lebzelter
          \inst{1}
          \and
          P.R. Wood\inst{2}
          \and
          K.H. Hinkle\inst{3}
          \and
          R.R. Joyce\inst{3}
          \and
          F.C. Fekel\inst{4}
          }

   \offprints{T. Lebzelter}

   \institute{Institute for Astronomy (IfA), University of Vienna,
              T\"urkenschanzstrasse 17, A-1180 Vienna, Austria\\
              \email{lebzelter@astro.univie.ac.at}
			\and Research School for Astronomy \& Astrophysics,
			Mount Stromlo Observatory, Weston, ACT 2611, Australia
			\and National Optical Astronomy Observatory\thanks{operated by the Association of
			Universities for Research in Astronomy, Inc. under cooperative agreement with the
			National Science Foundation}, PO Box 26732, Tucson, AZ 85726, USA
			\and Center of Excellence in Information Systems, Tennessee State University, 
			Nashville, TN 37203-3401, USA
                           }

   \date{Received ; accepted }

   \abstract{We present near infrared velocity curves for 12 long period variables (LPVs) in
   the globular cluster 47 Tuc (NGC 104).  New light curves are also presented for these variables. 
   Results are compared with   
the period-luminosity
   sequences occupied by the LPVs in the LMC.
Sequence C variables (fundamental mode pulsators) have larger velocity amplitudes
than sequence B variables (first overtone pulsators).
We show that, at similar luminosities, higher mass loss rates are associated with larger pulsation amplitudes.
One variable (V18) does not fit the normal period luminosity sequences and it has
an unusually large amount of circumstellar dust, suggesting that it has recently
undergone a thermal pulse on the AGB.
   Finally, we report the discovery of three new long period variable stars in the cluster core, all previously
   found to have a large infrared excess.
   }

   \maketitle
%
%________________________________________________________________

\section{Introduction}

In recent years, there have been significant advances in our understanding of
pulsation in long period variable stars (LPVs).  Particularly important is
the discovery of multiple period-luminosity relations for AGB stars in the LMC
and the interpretation of all but one of these relations in terms of radial
pulsation in low order modes (Wood et al.\,\cite{Wood99}).  Another advance is
the realization of the importance of pulsation in the production of mass loss
from red giants (e.g.  Wood \cite{Wood79}; Bowen \& Willson \cite{BW91}; H\"ofner et
al.\,\cite{Hoefner96}).

One problem with the observational study of pulsation and mass loss in red giants 
is that nearly all the stars studied so far are field stars.   This means that
we have no way of directly determining the initial or current stellar mass.
Any relations that are found for the field stars will be
broadened by the range in mass (age) and metallicity existing among the stars.
For Galactic stars, but not LMC stars, we have the additional problem that the luminosity (distance) is
generally poorly known.  As a result of these uncertainties, precise comparison of observed
periods, luminosities, amplitudes and mass loss rates with theoretical
calculations is complicated.
A way around these problems is to observe pulsating red giant stars
in star clusters where the giants have a common (initial) mass, composition
and distance.  We can then readily compare theoretical and observed 
period-luminosity relations, and we can see how the mass loss rate depends on
pulsation properties and quantities such as mass, luminosity and metallicity.

The driving of mass loss by pulsation is a complicated process.
Levitation of the stellar atmosphere caused by the pulsation
leads to the formation above the photosphere of a cool and dense
environment where dust grains form and grow efficiently. 
Radiation pressure on dust grains then combines with momentum in shock waves to 
drive the mass loss
(e.g. Wood \cite{Wood79}; Bowen \& Willson
\cite{BW91}; H\"ofner et al.\,\cite{Hoefner96}).  
It is found that larger pulsation velocity amplitudes and higher
luminosities drive a higher mass loss rate.  
One of the main purposes of this study is to
observe the pulsation velocity amplitudes in a good sample of red
variables in a star cluster so that a comparison of mass loss rate with pulsation velocity
amplitude (and luminosity) can be made.

The pulsation velocities of LPVs are best measured in the
near infrared since optical absorption lines measure velocities in the outer layers only.  These
optical velocities are nearly always directed inward relative to the stellar center-of-mass
(e.g. Wood \cite{Wood79}).  The near-infrared lines reveal the deeper, larger amplitude pulsation.
Monitoring of the velocity variations in field variables
using infrared (1.6 $\mu$m) CO lines has been carried out by Hinkle (\cite{Hinkle78}), Hinkle et al.\,({\cite{HHR82}; \cite{HLS97}) and
Lebzelter et al.\,(\cite{LKH00}).  These observations revealed velocity curves with 
amplitudes between 3 and 30\,km/s. The velocity curves of Miras are sawtooth-shaped and of large amplitude, with
phases of line doubling around maximum light.  The typical velocity curves
of semiregular variables (SRVs) are of smaller amplitude than in Miras and they are 
rather sinusoidal (i.e. continuous with no line doubling), although
they can reflect the partly irregular behaviour seen in the light curves.
A summary is given in Lebzelter \& Hinkle (\cite{LH00}).

The
globular cluster 47 Tuc (NGC 104) was chosen for this study since
it contains the richest known
collection of long period variables (LPVs).  Properties of this cluster from the literature are
summarized in Sect.\,\ref{param}. Along
its giant branch, four Miras and more than 10 semiregular
and irregular variables have been detected (Sawyer-Hogg \cite{Sawyer73}; Lloyd-Evans \cite{LE74}). Several infrared 
searches for indications of mass loss
from giant branch stars have also been reported (see Sect.\,\ref{massloss}).

\section{47 Tuc and its variables}\label{param}

47 Tuc is a prototype ``metal-rich'' globular cluster as well as being one of the closest.  
Recently, Gratton et al. (\cite{Gratton03}) presented a very detailed study on the fundamental
parameters of 47 Tuc. They derive a metallicity of [Fe/H]$=-$0.66$\pm$0.04 on the Carretta \& Gratton (\cite{CG97})
scale, which is in agreement with the findings of several other groups (see references in Gratton et al.). A metallicity
based on the Zinn \& West (\cite{ZW84}) scale is given by e.g. Briley et al. (\cite{Briley95}) with [Fe/H]=$-$0.76.

An accurate distance to the cluster is still a matter of some debate. Values in the literature for
the distance modulus of 47\,Tuc scatter by about $\pm$0.2 mag (for an overview see Gratton et al.\,\cite{Gratton03}).
Gratton et al. (\cite{Gratton03}) find
a distance modulus of \mbox{(m$-$M)}$_{V}$$=$13.50$\pm$0.08 from main sequence fitting. 
All studies agree that there is very little reddening towards 47 Tuc ($E(B-V)=$0.024$\pm$0.05).
Based on their estimates for the distance and metallicity Gratton et al. (\cite{Gratton03}) derived an age of 47 Tuc of 11.2$\pm$1.1\,Gyr. 
Hesser et al.\,(\cite{Hesser87}) derived a turnoff mass of 0.9\,M$_{\sun}$, but used an age estimate of
13.5\,Gyr. However, with the age from Gratton et al. and the more recent theoretical isochrones from Bertelli et 
al.\,(\cite{Bertelli94}) we again come to a turnoff mass between 0.86 and 0.9\,M$_{\sun}$.

A large number
of photometric measurements in the blue, visual and near- and
mid-infrared (Lee \cite{Lee77}; Frogel \cite{Frogel83}; Montegriffo et al.\,\cite{Montegriffo95}; Origlia
et al.\,\cite{Origlia97}; Ramdani \& Jorissen \cite{RJ01}) establish a very accurate color-magnitude diagram of
this cluster showing a well defined AGB. Luminosities and surface
temperatures have been derived for the AGB stars (e.g. Whitelock \cite{Whitelock86}). 

Light curves of the LPVs V1 to V8\footnote{Throughout this paper we use the nomenclature 
from the catalogue of globular cluster variables by Clement et al. (\cite{Clement01}).} 
were measured by Arp et al.\,(\cite{Arp63}). Several additional variables were detected
by Lloyd-Evans \& Menzies (\cite{LM73}}). A further study, including the variables V3 to V7, 
V11(=W12), V13 and V18, was presented by Fox (\cite{Fox82}). The light variability of these
stars will be further discussed in Sect.\,\ref{lightcurves}.

\section{Measurements of mass loss in 47 Tuc}\label{massloss}

There have been many attempts to measure the mass loss from AGB stars
in globular clusters. 
Mass loss at the end of the red giant branch (RGB, also called the first giant branch or FGB) phase
has already been proposed to explain the observed gaps in the horizontal
branch of globular clusters (Soker et al.\,\cite{soker01}; Schr\"oder \& Sedlmayr \cite{ss01}).
Cohen (\cite{cohen76}) proposed the existence of circumstellar shells (and hence mass loss)
around first giant branch stars in globular clusters to explain emission components
in the H$\alpha$ line. Bates et al.\,(\cite{BCK90}) summarized different studies on
H$\alpha$ lines of globular cluster stars and found a further indication for
circumstellar shells from the Na\,D line profiles. Lyons et al.\,(\cite{lyons96})
discussed mass motions in the atmospheres of 63 red giants from five different globular clusters
with the help of H$\alpha$ and Na\,D lines. Core shifts of these lines indicate
mass flow. The lower luminosity limit for outflow from H$\alpha$ and Na\,D lines is log(L/L$_{\sun}$)$=$2.5
and 2.9, respectively. Only part of the red giants showed an outflow. 
Standard deviations in the final masses of white dwarfs in globular
clusters of the order of 0.1\,mag may indicate a stochastic nature of the mass loss
(e.g.\,see the discussion by Alves et al.\,\cite{ABL00}).
%}

Not only is mass loss known to be essential in
stellar evolution but the mass loss from red giants in globular
clusters should result in intracluster gas and dust (Evans et al.\,\cite{Evans03}). 
Searches for the intracluster gas have
been limited in their results. The low metallicity no doubt plays a role in reducing 
silicate emission, making dust detection difficult (Frogel \& Elias \cite{FE88};
Helling et al.\,\cite{Helling02}). However, findings from AGB stars in
the Large Magellanic Cloud (LMC) confirm that low metallicity stars reach high mass loss rates
(e.g.\,Wood et al.\,\cite{Wood92}; van Loon et al.\,\cite{vanloon99}).

A few recent papers provide more detailed 
information on circumstellar shells around AGB stars in 47 Tuc. 
One of the most definitive results is by Ramdani \& Jorissen
(\cite{RJ01}) who used the ISO satellite to measure the mid-infrared 
emission around
6 AGB variables in 47 Tuc. Their findings, however, highlight the difficulty of
correlating pulsational properties with mass loss. Half of the variables,
namely V5, V7 and V15, show no or only marginal 12$\mu$m-excess while the other
three stars, V3, V11 and V18, do have a detectable excess flux at 12$\mu$m.
The excess of V11 is rather small. V3 is a Mira and would be expected to show an excess.
But the largest infrared excess was measured for V18, an irregular
AGB variable of only modest luminosity. 
Origlia et al. (\cite{Origlia02}) suggested that the mass loss from red giant
stars is episodic. They found an IR excess for V8 and no excess for V21
(see also Sect.\,\ref{massloss2}).

Overall, some indication of the presence or absence of circumstellar material
exists for 12 AGB variables in 47 Tuc. 
A substantial infrared excess in V3, V11 and V18 was detected by
Gillett et al. (\cite{Gillett88}), but these authors did not find indications for an infrared excess in V5 and V13.
An excess V-band polarization was found in V1 to V4, V6, V8 and V11 by Forte et al. (\cite{Forte02}) 
indicating the existence of circumstellar material around these stars.
Glass \& Feast (\cite{GF73}) reported an $L-$band excess in V3.
Frogel \& Elias (\cite{FE88}) detected a 10$\mu$m excess in the four cluster variables
V1, V2, V3 and V4. They calculated total mass loss rates for these objects ranging between
4.6 and 12.3x10$^{-6}$\,M$_{\sun}$yr$^{-1}$. A 10$\mu$m excess was also detected in V1 and V3
by Origlia et al. (\cite{Origlia97}). However their
estimated dust mass loss rates are about ten times less than those of Frogel \& Elias (\cite{FE88}).

\section{Observations and results}

We selected 12 long period variables in the globular cluster \mbox{47 Tuc} for which a period or at least some information 
on the variability type was given in the
catalogue of variable stars in globular clusters published by Clement et al. (\cite{Clement01}). 
These stars are moderately bright infrared sources, with
K$\sim$6-8. Table \ref{t:sample} summarizes the properties of the stars.

\begin{figure}
   \centering
   \resizebox{\hsize}{!}{\includegraphics{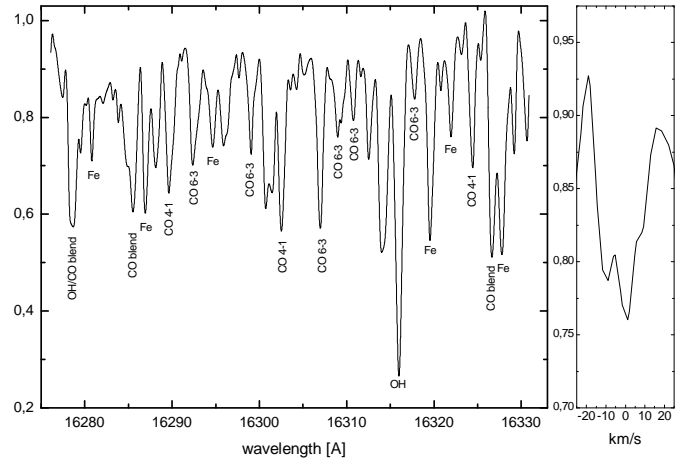}}
   \caption{Left panel: Example spectrum of the star V11. Several prominent features are identified.
   Right panel: Example of line doubling (CO 4-1, V3).}
   \label{specex}
   \end{figure}

Time series of infrared spectra in the $H$ band were obtained in 2001 and 2002 with the 74inch
telescope at Mount Stromlo Observatory, Australia. The NICMASS detector, already successfully used
for a preceding program at Kitt Peak (Joyce et al. \cite{Joyce98}), was used at the Coud\'{e} focus of the telescope.
The standard infrared observation technique was used. Spectra of each star were obtained at two different slit positions
to allow sky subtraction. 
Resolution was set to R$=$37000. We achieved a S/N ratio of 30 or better. An example spectrum is shown in the left part of
Fig.\,\ref{specex} with an identification of some relevant spectral features.
The spectral range covered a number of second overtone CO lines,
some OH lines, and a few metallic lines.
When 47\,Tuc was visible at Mount Stromlo spectra were obtained approximately once a month. 
The observing program had an unexpected end when Mount Stromlo Observatory was destroyed
by a bush fire in early 2003. Nevertheless time coverage and sampling of the 47\,Tuc AGB variables,
combined with a few additional spectra taken in the same wavelength region with the PHOENIX spectrograph 
at Gemini South, is sufficient for the investigation presented here.

The bright stars $\alpha$ Cet and $\delta$ Oph have been used as primary velocity
standards (Udry et al.~\cite{Udry99}). Velocities of the variables were determined by a cross correlation technique,
using the IRAF task fxcor. Typical velocity uncertainties, determined from multiple
observations of some stars in the same or consecutive nights, were found to be
$\sim$0.4 kms$^{-1}$.

To determine
proper phases for the velocity curve, photometric measurements of all variables in our sample except V5
(see below) were done as well. 
Data in the blue and red MACHO filters were obtained
with the 50inch telescope at Mount Stromlo (also destroyed in the January 2003 fires). $V$ and $I$ measurements
were obtained with ANDICAM at the YALO telescope (before the MACHO dataset), with a 
few additional measurements from ANDICAM at the CTIO 1.3m telescope (after the MACHO dataset). 
The different data sets have been combined
using transforms based on about 40 nonvariable cluster stars ranging in $V-I$ between 0.9 and 2.1. 
The MACHO blue filter has a mean wavelength close to that of the $V$ filter
so MACHO blue magnitudes can be reliably transformed to $V$.  Long time series
are therefore available in the $V$ filter. No obvious phase shift relative to the other
filters was noticed, but a conclusive result cannot be drawn from our data.
From the $V$ light curves we determined
the time of the light maxima, phase zero.
%} 

\subsection{Light curves}\label{lightcurves}

Light curves for most of the variables are shown in the upper panels of Figs.\,\ref{f:v1} to \ref{f:v21}. 
Periods derived from our measurements were in most cases in good 
agreement with the values from
the literature. 
In our data V4 shows a 
main period of $\sim$170\,d in agreement with
the value of 165\,d from Arp et al.\,(\cite{Arp63}).
Fox (\cite{Fox82}) found a period of 82 days for V4 and suggested that this star switched between two modes.  
This type of multimode 
behaviour has subsequently
been found to be common in the light curves of LPVs (e.g. Wood et al.\,\cite{Wood99}).
For V5, we unfortunately do not have sufficient usable data as this star was located in a bad area of the CCD. The Sawyer-Hogg value
(\cite{Sawyer73}) is 60 days, but different values can be found in the literature, ranging between 29 and
70\,days (Fox \cite{Fox82}). From our velocity data we rather favour a period of 50 days.
The short periods of 52 and 100 days reported for V11 could not be confirmed in our data. Instead, the star shows some long
period variation lasting more than 200 days (Fig.\,\ref{f:v11}). Compared to the data presented by Fox (\cite{Fox82})
the amplitude of the variation is smaller.  This star appears to be another multimode pulsator.  For V13, there is a hint of the 
catalogued $\sim$40\,d period in our light curve,
but a period of $\sim$90\,d may also be present.  Additionally, our light curve suggests a 
long period (Fig.\,\ref{f:v13}) about 10 times the short period.  This star appears to be yet another multimode pulsator.

V18 was reported variable both
in the catalogue of Sawyer-Hogg (\cite{Sawyer73}) and by Lloyd Evans (\cite{LE74}), while Fox (\cite{Fox82})
found no variability in this star. We found photometric variability with an amplitude  of about
0.2 mag in this object (Fig.\,\ref{f:v18}). Our 
short light curve 
would suggest a period of about 83 days. 
For V21, no period determination existed in the literature. Our data set
(see Fig.~\ref{f:v21}) shows a period of 76 days.

\begin{table}
\caption{Properties of LPVs in 47\,Tuc.}\label{t:sample}
\begin{tabular}{lcccccc}
\hline 
Name & $\alpha$ (2000) & $\delta$ (2000) & $J$ & $K$ & P [d] \\
\hline 
\noalign{\smallskip}
V1  & 00 24 12.4 & $-$72 06 39 & 7.45 & 6.21 & 221    \\
V2  & 00 24 18.4 & $-$72 07 59 & 7.52 & 6.29 & 203    \\
V3  & 00 25 15.9 & $-$72 03 54 & 7.49 & 6.27 & 192    \\
V4  & 00 24 00.3 & $-$72 07 26 & 7.87 & 6.69 & 165,82 \\
V5  & 00 25 03.7 & $-$72 09 31 & 8.65 & 7.47 & 50     \\
V6  & 00 24 25.5 & $-$72 06 30 & 8.54 & 7.43 & 48     \\
V7  & 00 25 20.6 & $-$72 06 40 & 8.18 & 6.97 & 52     \\
V8  & 00 24 08.3 & $-$72 03 54 & 7.94 & 6.70 & 155    \\
V11 & 00 25 09.0 & $-$72 02 17 & 7.91 & 6.71 & 52,160?\\
V13 & 00 22 58.3 & $-$72 06 56 & 8.79 & 7.70 & 40     \\
V18 & 00 25 09.2 & $-$72 02 39 & 8.59 & 7.47 & 83?    \\
V21 & 00 23 50.1 & $-$72 05 50 & 8.07 & 6.78 & 76    \\
\noalign{\smallskip}
\hline
\end{tabular}
Notes: 
For V18 and V21 both period determinations are from this study. For V1
we suggest a period of 221 days from our data 
instead of 212 days listed in the literature. For V5 we favour a
period of 50 days (instead of 60 days listed in the Clement et al.~catalogue). 
For V11 see text. All other periods are from the catalogue of Clement et al.~(\cite{Clement01}).
The $J$
and $K$ magnitudes are averages of the maximum and minimum observed values
from the following sources: the 2MASS Point Source Catalog; Fox (\cite{Fox82});
Menzies \& Whitelock (\cite{mw85}); and Frogel et al. (\cite{fpc81}).  All magnitudes have been
converted to the AAO system using conversions in Allen \& Cragg (\cite{allen83})
and Carpenter (\cite{car04}).
\end{table}

\subsection{Velocity curves}

Velocity curves were determined for all 12 stars of our sample. All stars show at least some velocity variability 
above our detection threshold. 
Line doubling was detected in three stars of our sample (Tab.\,\ref{t:results}).
An example of line doubling is shown in the right part of Fig.\,\ref{specex}.

\begin{figure}
   \centering
   \resizebox{\hsize}{!}{\includegraphics{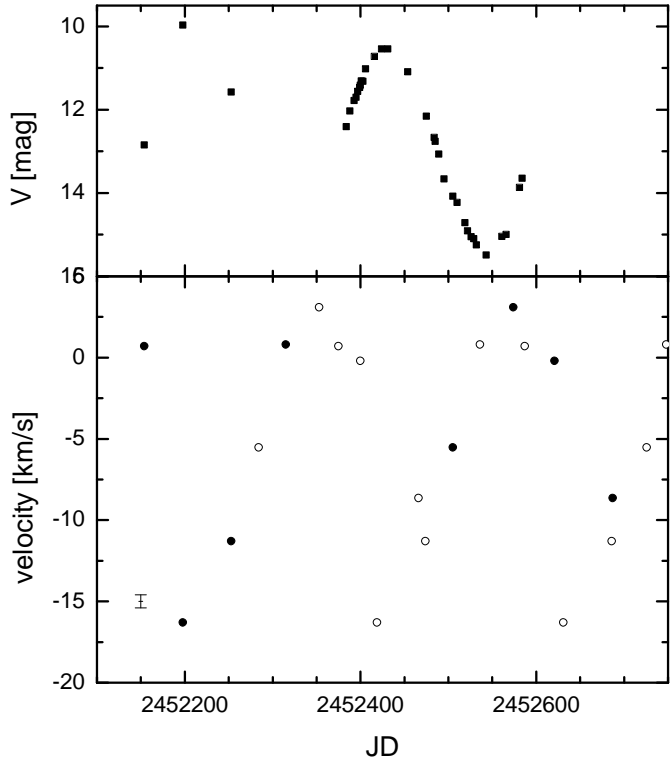}}
   \caption{Light (upper panel) and velocity variations (lower panel) for V1. 
   Filled symbols in the lower panel indicate individual velocity measurements.
   For a better illustration of the velocity change
   data are repeated shifted by an integral number of periods forward and backward in time (open symbols). 
   The period listed in Table 1 is used. The typical error bar for
   the velocity data is indicated.}
   \label{f:v1}
   \end{figure}

\begin{figure}
   \centering
   \resizebox{\hsize}{!}{\includegraphics{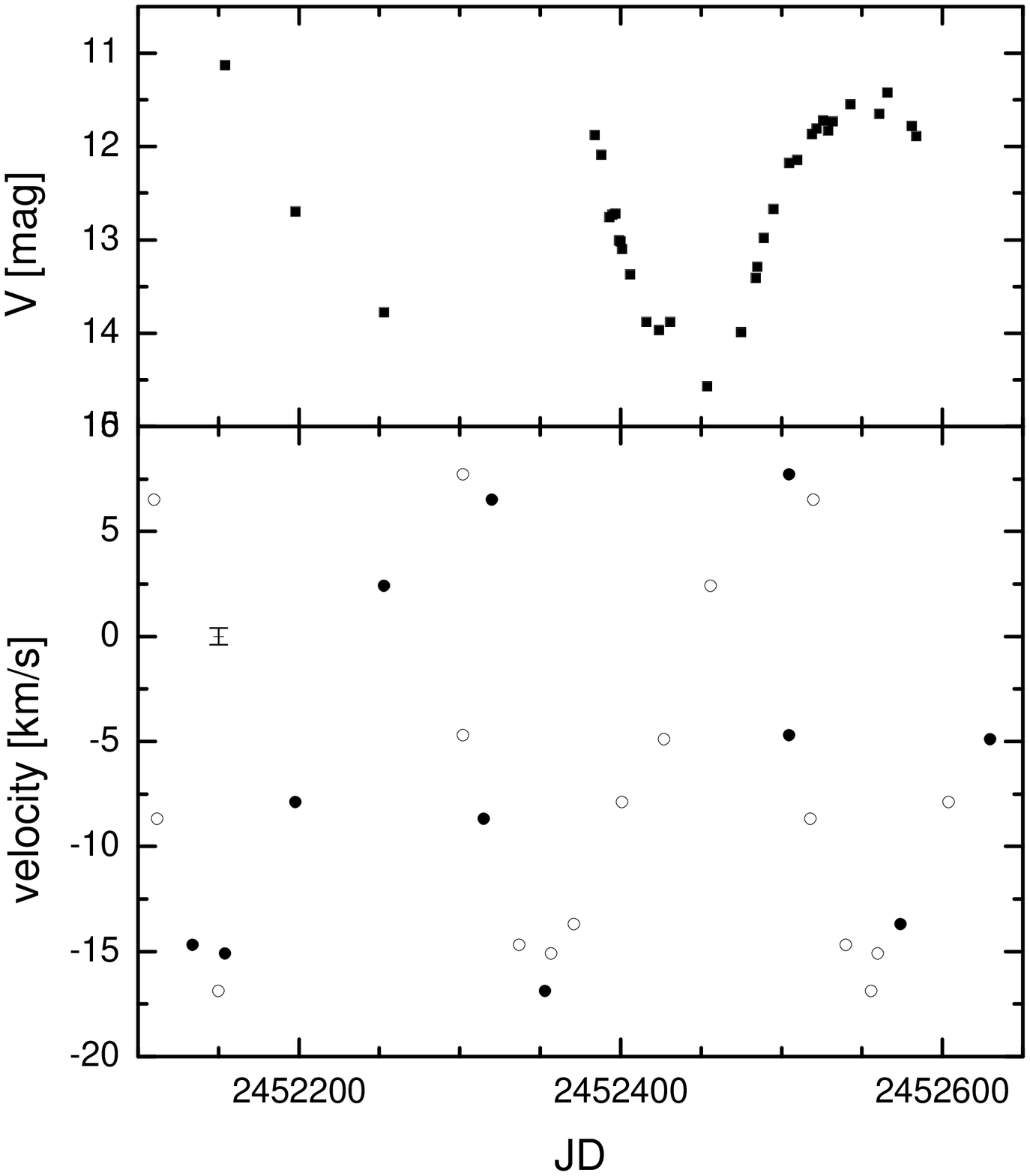}}
   \caption{Same as Fig.\,\ref{f:v1} for V2.}
   \label{f:v2}
    \end{figure}

\begin{figure}
   \centering
   \resizebox{\hsize}{!}{\includegraphics{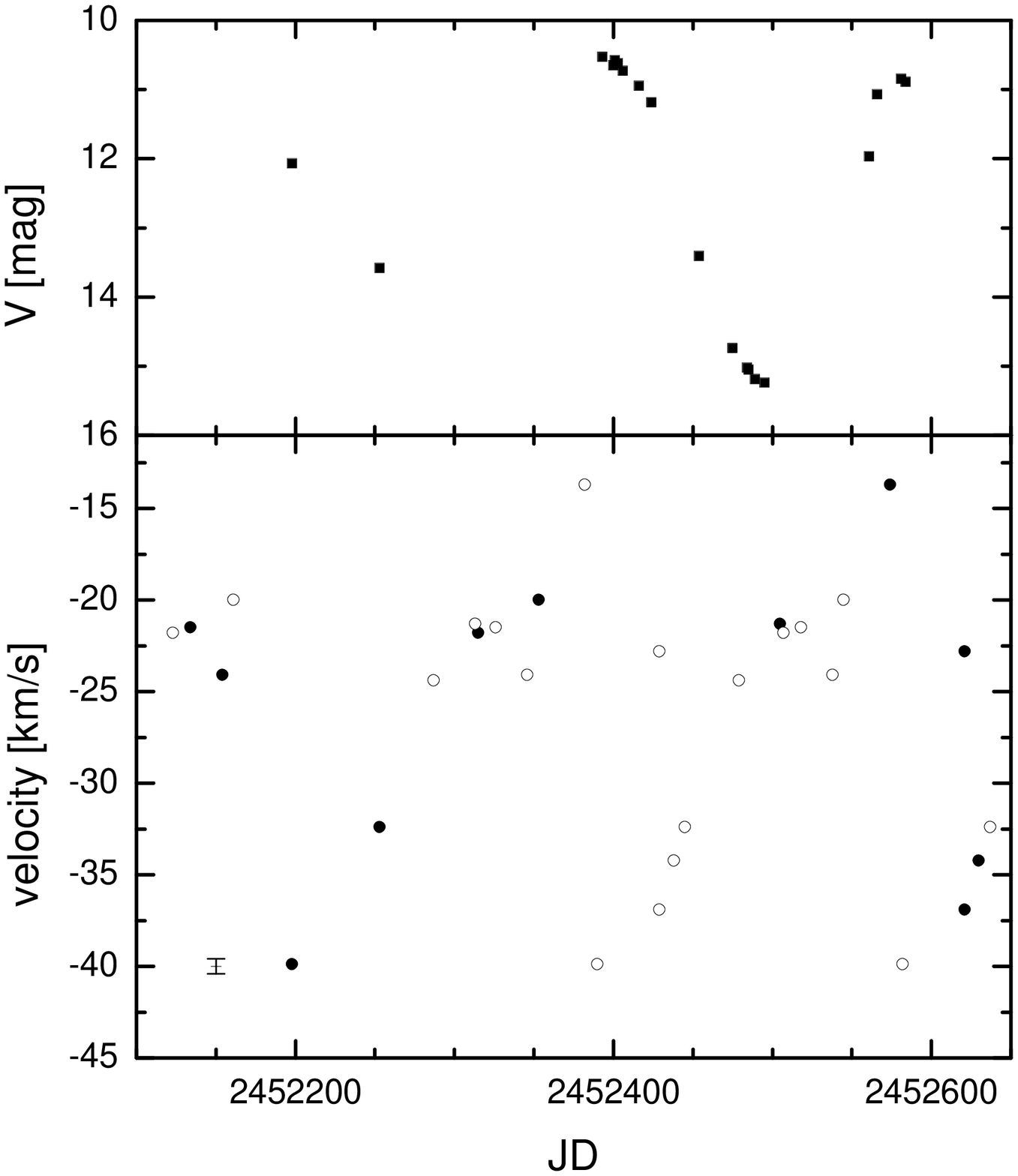}}
   \caption{Same as Fig.\,\ref{f:v1} for V3.}
   \label{f:v3}
    \end{figure}

\begin{figure}
   \centering
   \resizebox{\hsize}{!}{\includegraphics{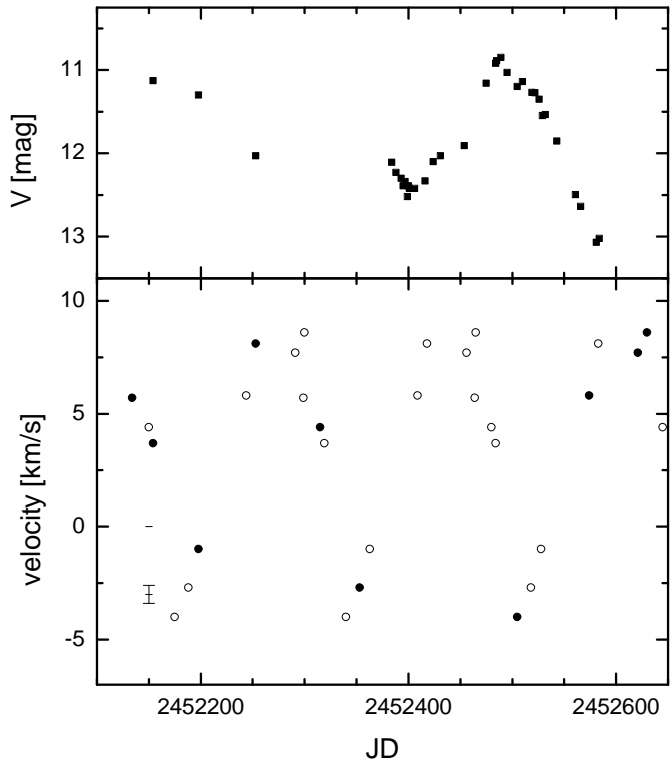}}
   \caption{Same as Fig.\,\ref{f:v1} for V4.}
   \label{f:v4}
    \end{figure}

\begin{figure}
   \centering
   \resizebox{\hsize}{!}{\includegraphics{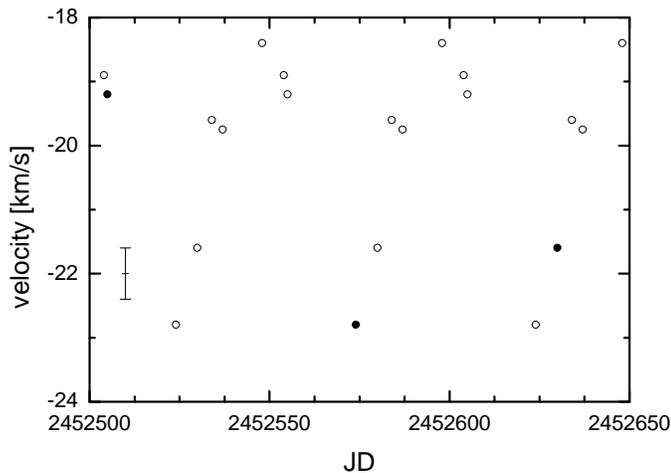}}
   \caption{Velocity change of V5. Only part of the time of monitoring is shown. Symbols as
   in the lower panel of Fig.\,\ref{f:v1}. No parallel light curve data exist for this star.
   A typical error bar for the velocity data is indicated.}
   \label{f:v5}
    \end{figure}

\begin{figure}
   \centering
   \resizebox{\hsize}{!}{\includegraphics{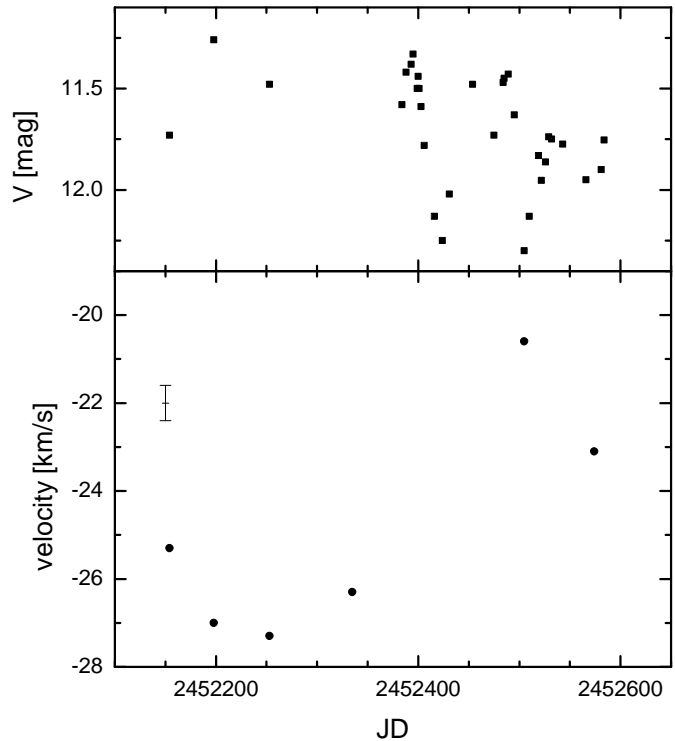}}
   \caption{Light (upper panel) and velocity (lower panel) variations for V6.
   A typical error bar for the velocity data is indicated.}
   \label{f:v6}
    \end{figure}

\begin{figure}
   \centering
   \resizebox{\hsize}{!}{\includegraphics{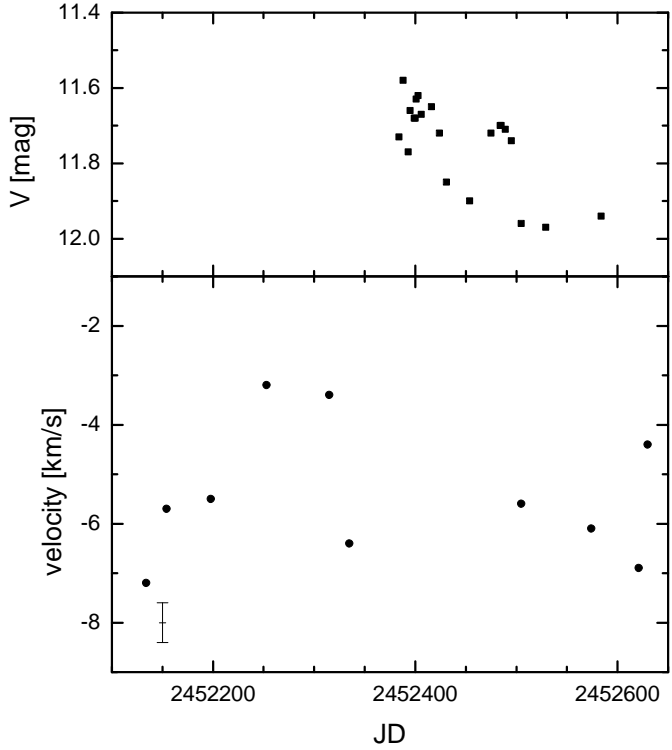}}
   \caption{Same as Fig.\,\ref{f:v6} for V7.}
   \label{f:v7}
    \end{figure}

\begin{figure}
   \centering
   \resizebox{\hsize}{!}{\includegraphics{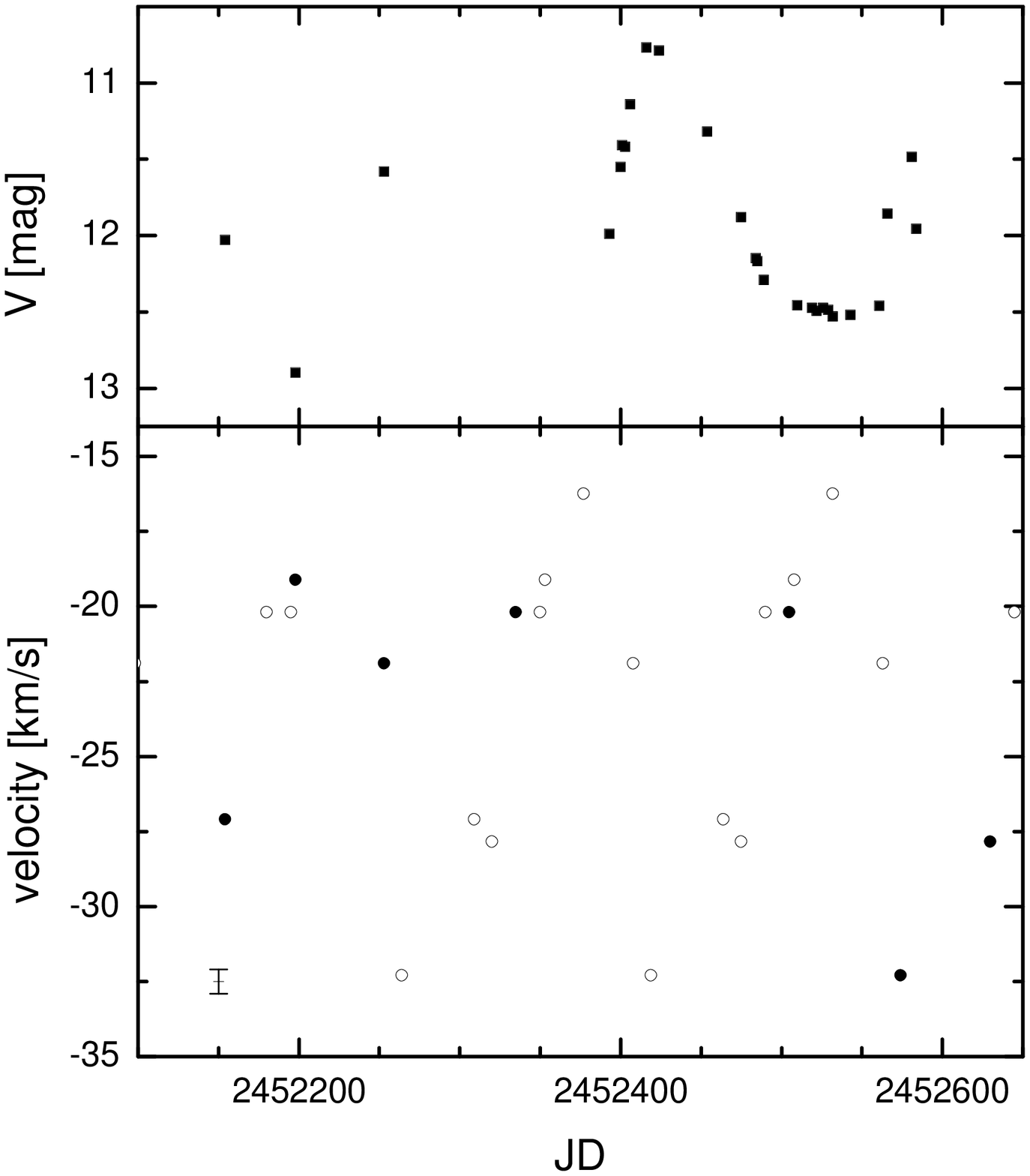}}
   \caption{Same as Fig.\,\ref{f:v1} for V8.}
   \label{f:v8}
    \end{figure}

\begin{figure}
   \centering
   \resizebox{\hsize}{!}{\includegraphics{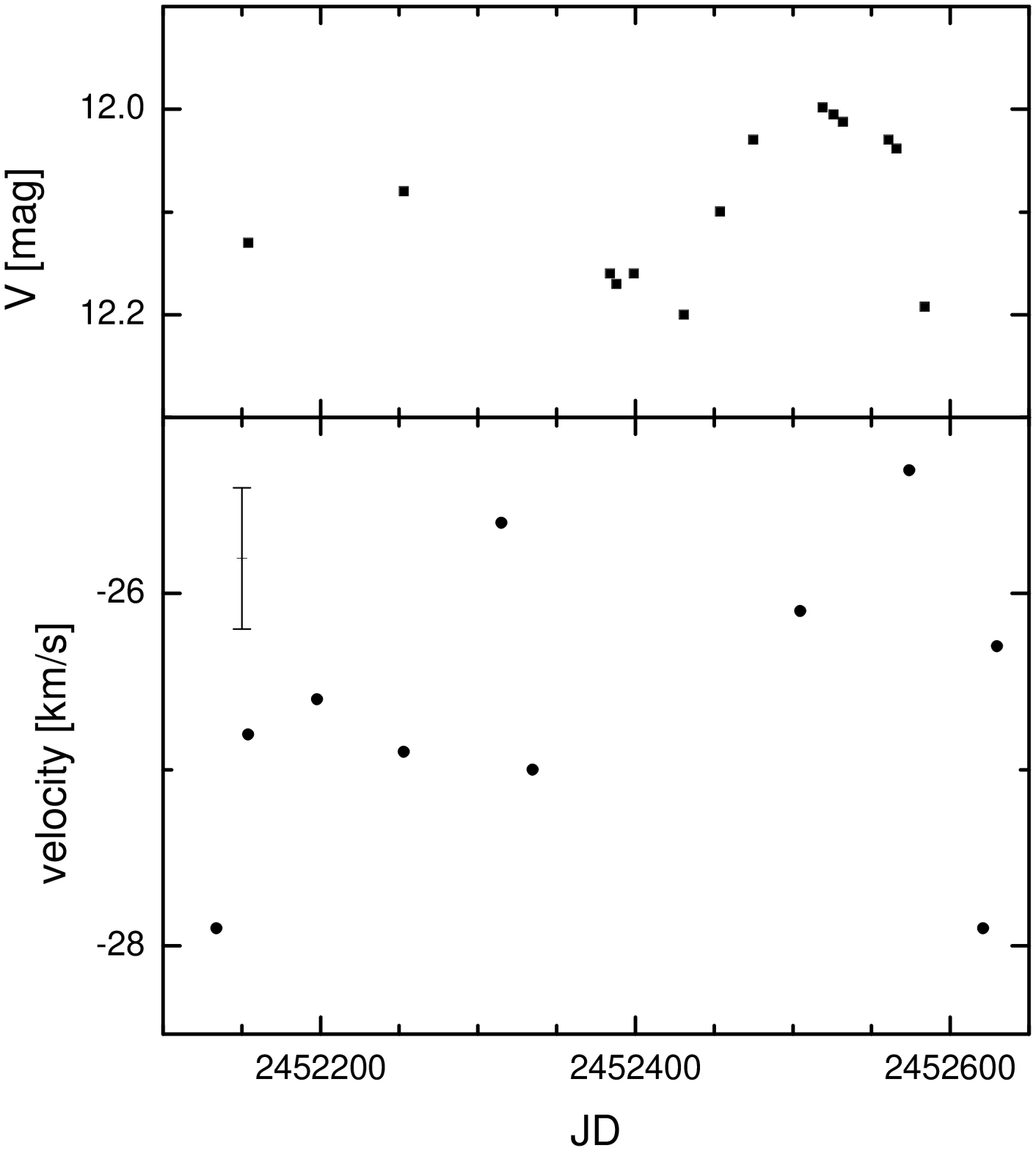}}
   \caption{Same as Fig.\,\ref{f:v6} for V11.}
   \label{f:v11}
    \end{figure}

\begin{figure}
   \centering
   \resizebox{\hsize}{!}{\includegraphics{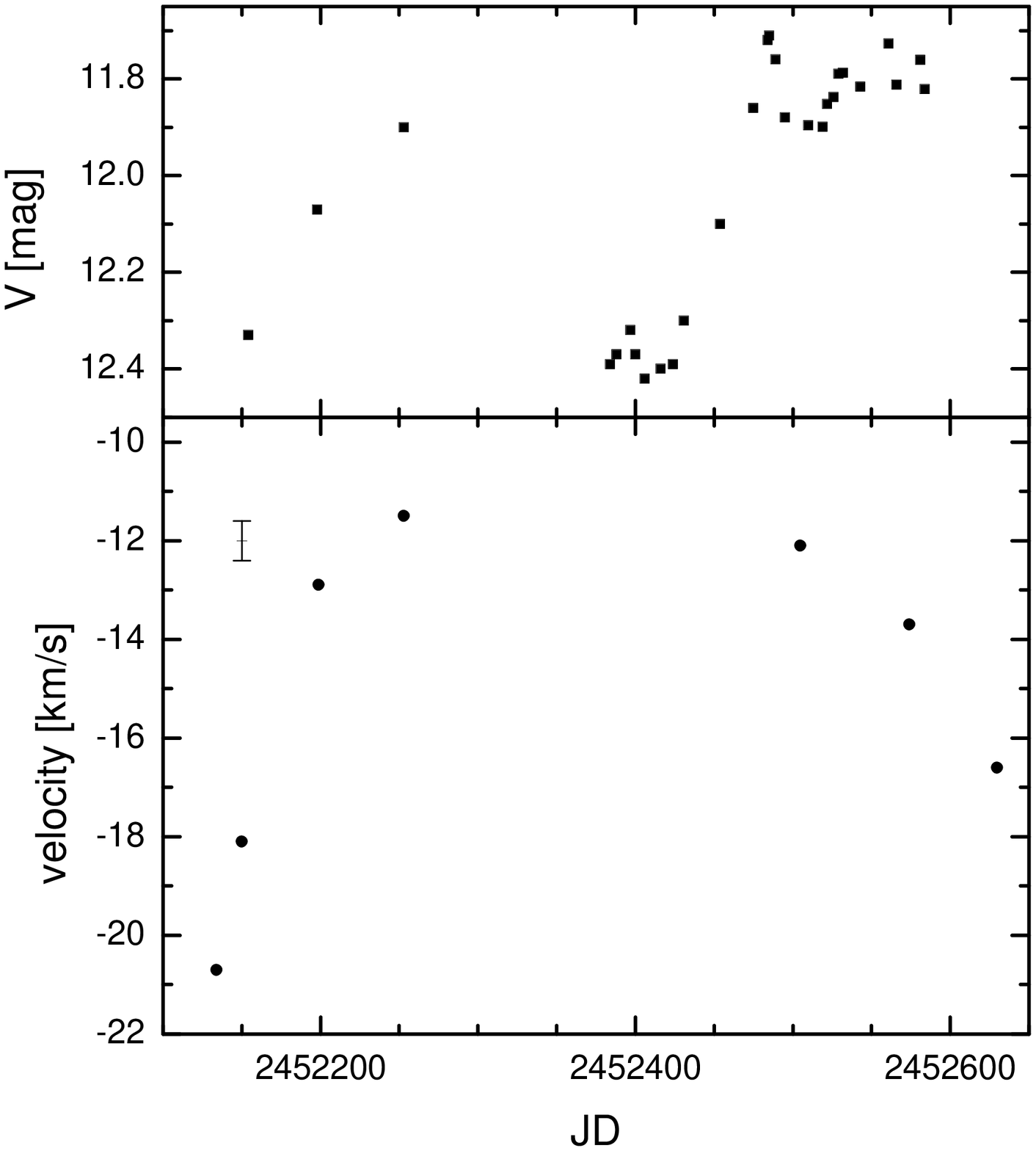}}
   \caption{Same as Fig.\,\ref{f:v6} for V13.}
   \label{f:v13}
    \end{figure}

\begin{figure}
   \centering
   \resizebox{\hsize}{!}{\includegraphics{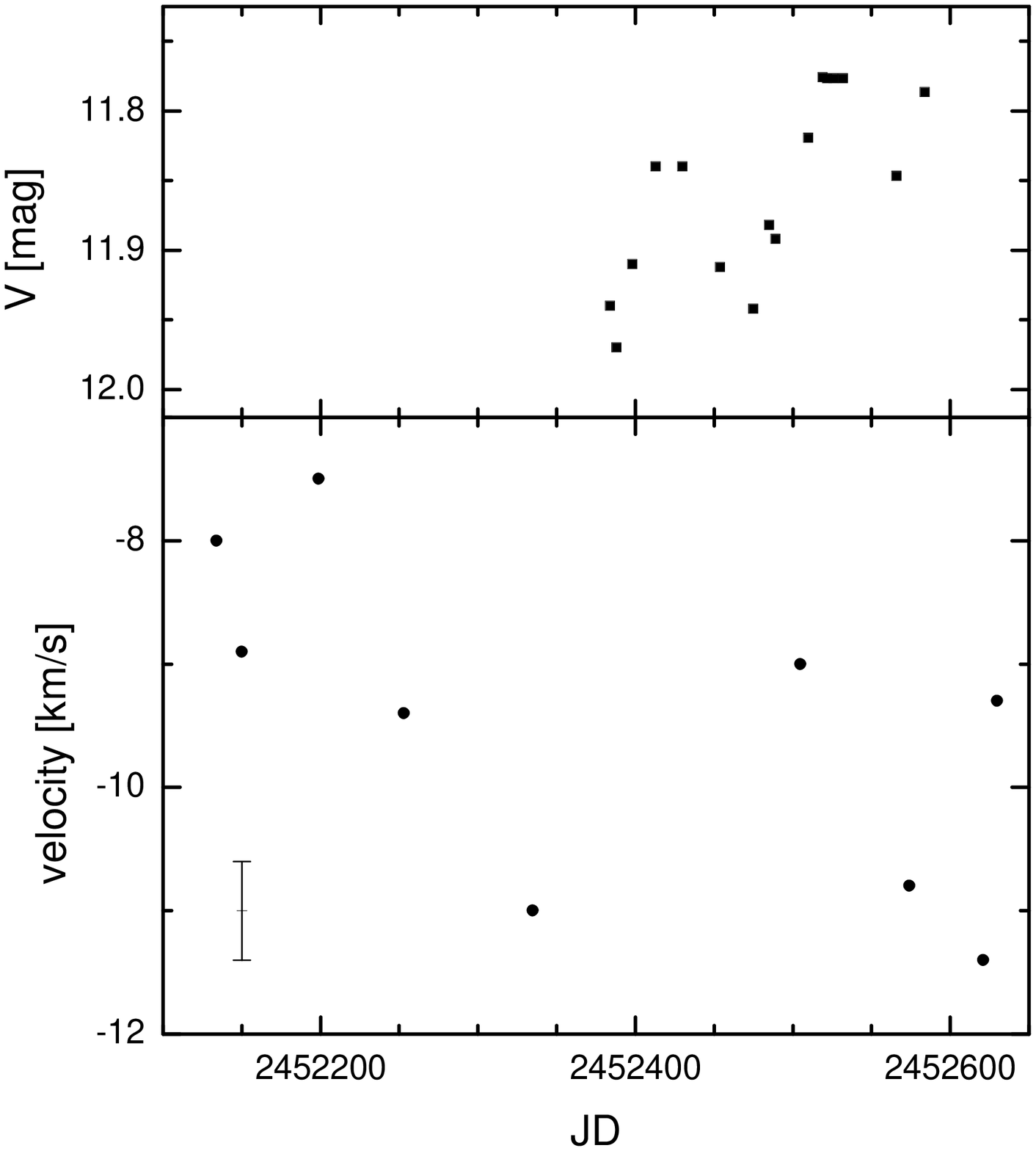}}
   \caption{Same as Fig.\,\ref{f:v6} for V18.}
   \label{f:v18}
    \end{figure}

\begin{figure}
   \centering
   \resizebox{\hsize}{!}{\includegraphics{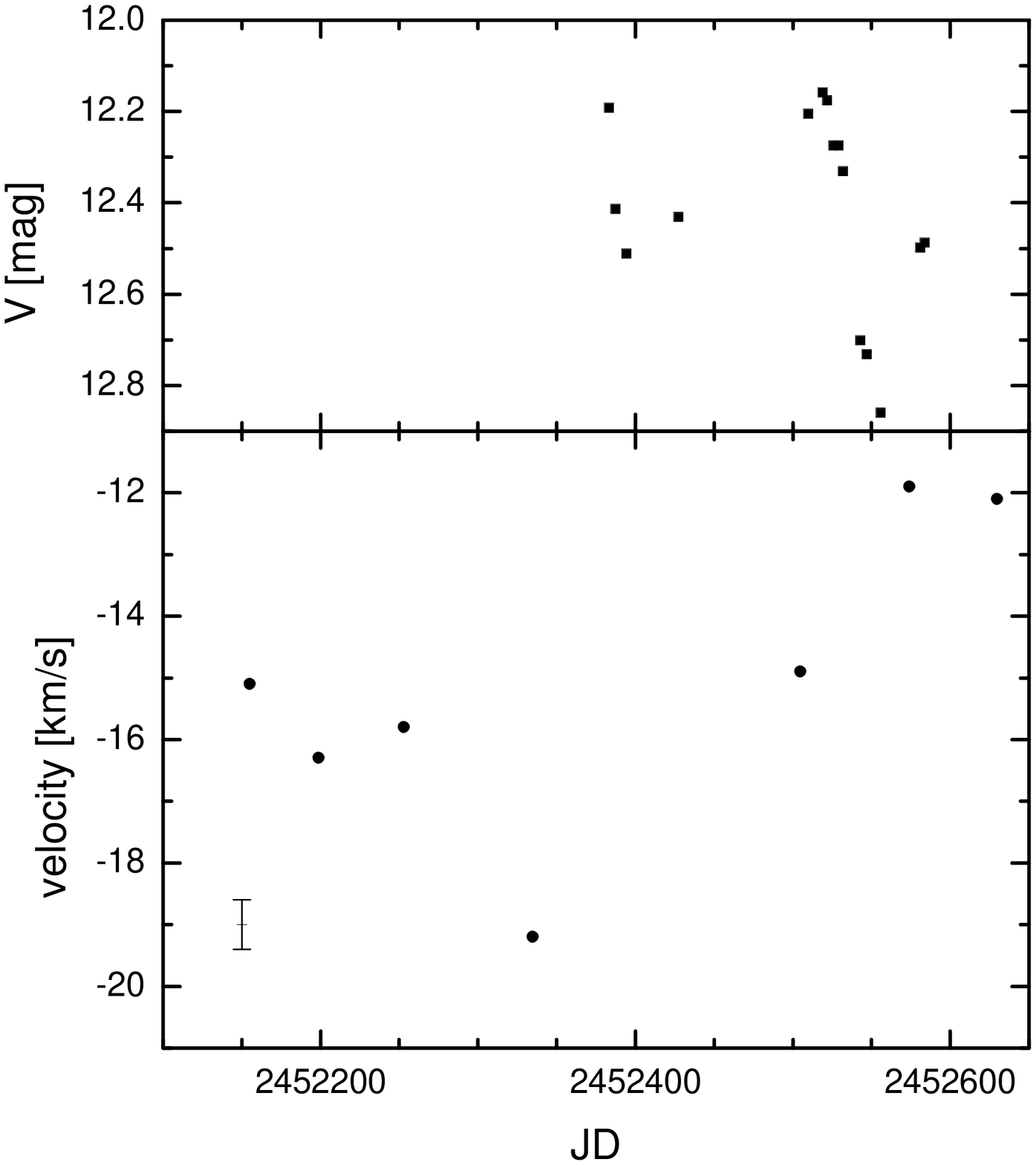}}
   \caption{Same as Fig.\,\ref{f:v6} for V21.}
   \label{f:v21}
    \end{figure}

The velocity curves we derived are shown in the lower parts of
Figs. \ref{f:v1} to \ref{f:v21} together with the corresponding light change. 
The velocity curves can be roughly separated into three
groups: V1, V2, V3, V4, and V8 show velocity curves very similar in shape to those typical of Miras
found in the solar neighborhood (Lebzelter \& Hinkle 2001). The velocity amplitude of V1, V2 and
V3 are all similar to nearby Miras, too. The other two stars show a similar shape but a clearly
smaller amplitude. While we confirm the Mira-like nature of these five variables (e.g. noted by
Whitelock \cite{Whitelock86}), we note that the smaller velocity amplitude of the latter two stars
is also accompanied by a smaller light amplitude.
The second group of variables, consisting of V5, V6 and V7 is comparable to local semiregular variables (Lebzelter \& Hinkle 2001).
Their amplitudes are much smaller than in the first group. V5 shows a very regular light change, while the other two
objects are obviously not strictly periodic, in agreement with the semiregular nature of their light curves. The third group
consists of V11, V13, V18, and V21. In all four of these stars, there is no obvious correlation between the light
change and the velocity variations. In the case of V13, it cannot be ruled out that variations occur on a time scale similar
to the short or even the long period (see above). For V11 our observations do not allow a clear picture of the light change, so a correlation
is not possible. On the other hand, V18 and V21 do show a rather well defined light change, but velocity variability
occurs on a different time scale. The reason for that is not clear and longer spectroscopic and photometric time series will
be needed to understand this phenomenon.
Table \ref{t:results} summarizes the results on the velocity variations in the 47\,Tuc AGB variables. The table gives the
total velocity amplitude, the characteristics of the velocity curve and comments on the occurrence of line doubling.

\begin{table}
\caption{Data on the AGB variables in 47\,Tuc.}\label{t:results}
\begin{tabular}{lccc}
\hline 
\noalign{\smallskip}
Name & velocity amplitude & variability & comments\\
 & [km/s] & type & \\
\hline 
\noalign{\smallskip}
V1 & 20 & Mira-like & line doubling\\
V2 & 23 & Mira-like & line doubling\\
V3 & 22 & Mira-like & line doubling\\
V4 & 18 & Mira-like & double-period\\
V5 & 8 & regular &\\
V6 & 7 & semiregular &\\
V7 & 4 & semiregular &\\
V8 & 16 & Mira-like &\\
V11 & 4 & irregular &\\
V13 & 12 & long period & double-period\\
V18 & 5 & irregular &\\
V21 & 7 & long period &\\
\noalign{\smallskip}
\hline
\end{tabular}
\end{table}

\section{Discussion}
\subsection{Pulsation and Stellar Evolution} \label{puls}

Observations from the Magellanic Clouds indicate that AGB variables are found on four distinct
period-luminosity relations (e.g. Wood \cite{Wood00}). Three of these sequences can be interpreted in terms of
stars pulsating in the fundamental mode and the first, second and third overtone modes\footnote{Sequence A is 
attributed to both second and third overtone mode pulsators.}. The fourth period-luminosity
relation still lacks a definite interpretation (Wood et al.\,\cite{Wood04}). For the Milky Way AGB
variables the absence of reliable distances limits the establishment of such a period-luminosity 
relation (e.g. Bedding \& Zijlstra
\cite{BZ98}). In the case of both the LMC and the Milky Way, 
the diversity of the stars in age, and therefore initial mass, complicates the
interpretation.

\begin{figure}
   \centering
   \resizebox{\hsize}{!}{\includegraphics{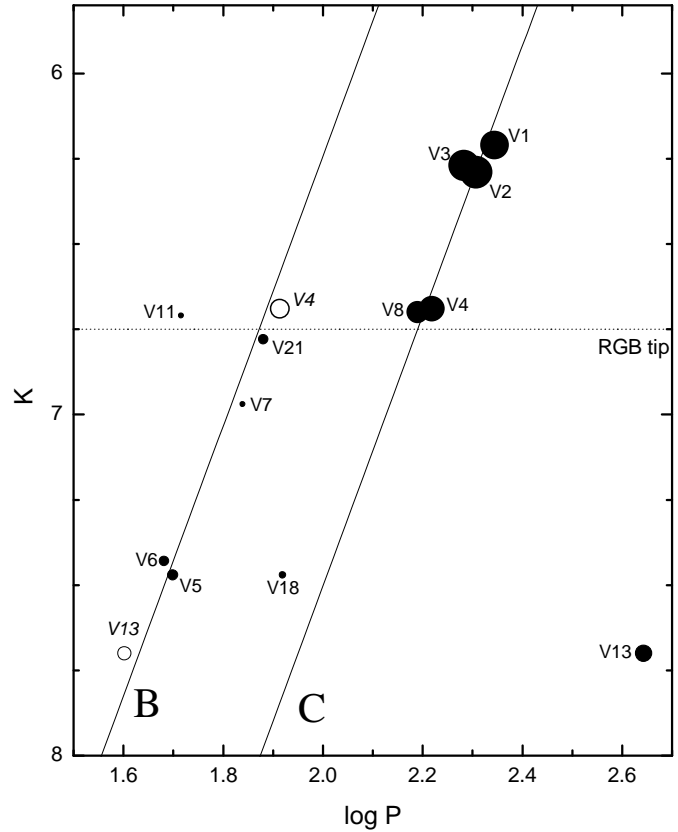}}
   \caption{logP vs. K diagram for the long period variables in 47\,Tuc. Symbol size
   denotes the velocity amplitude (ranging between 4 and 23\,kms$^{-1}$). Lines indicate
   the approximate location of sequences B and C found by Wood (\cite{Wood00}) for LMC
   long period variables shifted to the distance of 47\,Tuc. The dotted horizontal line
   marks the tip of the RGB according to Ferraro et al.\,(\cite{FMOF00}).
   Stars with possible multiple periods are shown twice.}
   \label{f:pld}
    \end{figure}

The AGBs of globular clusters are characterized by a much more homogeneous set of stellar parameters. The richness
of 47\,Tuc in AGB variables allows us to use this advantage for a discussion of stellar pulsation in relation to
luminosity, i.e. evolutionary stage. Period-luminosity diagrams for globular clusters have been constructed
by several authors (e.g.\,Feast et al.\,\cite{Feast02}). Typically, only a relation for the large amplitude, long period
variables ('Miras') was derived. This relation is in agreement with the sequence C found from LMC data
(Wood \cite{Wood00}). In Fig.\,\ref{f:pld} we placed
the AGB variables of 47\,Tuc studied in this paper into a logP-K-diagram. The approximate location of the P-K-relations B (first overtone)
and C (fundamental mode) determined in the LMC (Wood \cite{Wood00}) are indicated by lines. Relations were transformed from the
LMC to 47\,Tuc distance using a LMC distance module of 18.515 (Clementini et al.\,\cite{Clementini03}).  Stars with more
than one well established period in the literature or from our own light curve data 
are plotted twice and are marked accordingly in Table \ref{t:results}.

It can be seen that the variables
nicely follow these two relations. The good fit of the P-L-relations to the 47\,Tuc data after scaling from the LMC
is indeed remarkable. Obviously, the physical mechanism that determines the location of at least these two
sequences in the logP-K-diagram is the same for LMC and 47\,Tuc. Furthermore, the distance moduli
we used in this paper seem to be very well determined at least concerning the relative distance between LMC and
47\,Tuc. However, it has to be noted that the location of the sequences in the LMC is not
accurately determined, as there is a significant scatter (probably due to the larger scatter in stellar parameters and maybe
also the extension of the LMC) around the average locations used here. We therefore do not want to overinterpret our findings on that point.

There is a clear bipartition in our logP-K-diagram (Fig.\,\ref{f:pld}): At lower luminosities only sequence B is
populated (see below for the case of V18), no variables are known in 47\,Tuc that would fall on the lower part of
logP-K-relation C. It is very likely that the stars V5, V6, V7, and V13, which we find in this lower luminosity part
are probably in an earlier evolutionary state than the stars at higher luminosity (see below). 
In a first approach we assume that the luminosity of the stars in Fig.\,\ref{f:pld}
corresponds to the evolutionary status. 

At the highest luminosities
only sequence C is occupied, and the stars there are the Miras V1, V2, and V3.  Below these stars on sequence C lie V4 and V8.
Both stars are located close to the tip of the RGB. They have similar velocity amplitude and K brightness. V4 appears to switch modes between sequence C ($P=165$\,d) and sequence B ($P=82$\,d). From our data we have to conclude that
V4 is currently only on sequence C. 
V11 having approximately the
same K luminosity as V4 is another star suspected of showing mode switching (see Sect.~\ref{lightcurves}) 
located clearly off sequence B towards shorter periods (i.e. near sequence A). The period used in the plot is taken from Fox
(\cite{Fox82}). It is interesting that both stars with possible mode 
switching are found at similar K luminosity, namely close to the RGB tip.

For V13 we found indications
of a second period, and the star is therefore shown twice in Fig.\,\ref{f:pld}. 
The long period of V13 is located close to sequence D
from Wood (\cite{Wood00}). The nature of this period-luminosity relation is not yet understood (Wood et al.
\cite{Wood04}).

Symbol size in our P-K-diagram is linearly proportional to the measured velocity amplitude. 
In the case of V4 we stress again that the shorter period, marked with an open symbol in Fig.\,\ref{f:pld},
is clearly not dominating the light or velocity variations at the moment. The symbol size of
the V4 data point on sequence B may therefore be misleading as we have no indications that the velocity
amplitude would be the same if the 82 days period would dominate. According to our data the star
is currently monoperiodic.
It can be seen that the large amplitude variables are all
found along sequence C. Our results indicate further that the velocity amplitude is increasing along sequence C.
Velocity amplitudes along sequence B are much smaller and do not show a steady increase of velocity amplitude 
with luminosity. The large 
amplitude of V13 seems to be
associated with the long period ($\sim$400\,d) variation rather than the 40\,d variation of sequence B.

These results lead us to the following probable evolutionary scenario:
Stars evolve up the AGB from low luminosities. They pulsate first 
in the first overtone
(sequence B) then, at intermediate luminosities close to the RGB tip, they go through an interval of mode switching back and forth
between first overtone and fundamental mode and finally, at the highest luminosities, they remain pulsating
in the fundamental mode.  There is clearly an increase in both the light and velocity amplitude associated
with the switch from first overtone to fundamental. The large velocity amplitudes found at the tip of
sequence C are similar to what is expected from models for fundamental mode pulsation (e.g.\,Scholz \& Wood \cite{SW00}).

The role of V18 in this scenario is not clear. As noted in Section~\ref{massloss}, this star has one of the largest infrared excesses 
among the 47\,Tuc variables.
In Fig.\,\ref{f:pld} we find the star in between sequences B and C at a rather low luminosity. Its velocity amplitude is
rather small. However, locating this star on
the period axis is difficult as the star has shown periodic, nonperiodic and constant phases in the past.

Comparing our results with evolutionary models requires some adaption of this scenario: 
Due to luminosity variations during a Thermal Pulse the current location of a star in the logP-K-diagram may not correspond
to its evolutionary status along the AGB.
However, as was shown e.g.~by
Boothroyd \& Sackmann (\cite{BS88}) the probability to observe the star at luminosity values highly deviating from the
mean (and for the star's evolutionary status typical) value is comparatively low. This fact is even more expressed in stars
of 1\,M$_{\sun}$ or less where the interpulse time is very large compared to the duration of the pulse itself (Vassiliadis \& Wood \cite{Vass93}).
From the models of Vassiliadis \& Wood (\cite{Vass93}, see their Fig.\,13) for a star of 1\,M$_{\sun}$ and Z$=$0.004 
we would expect to
see about 1/6 of our sample in a phase strongly deviating from the mean global parameters of the star, in our case these would be 2 stars (if our sample consists of AGB stars only, see Sect.~\ref{agbrgb}). 
Most probably, these stars are expected to be found at a lower luminosity and a shorter period. It is therefore unlikely that
one of the stars on sequence C belongs to this group. Following these considerations we have to add to our scenario the possible occurrence of loops in the logP-K-diagram during the evolution 
up the AGB. V18 with its large IR excess and odd position in the logP-K-diagram may be an example of such a star.
However, it is unlikely that e.g.~the mode switch of V4 is an indication of such a loop because of the time scale of
the mode switch. Arp et al. (\cite{Arp63}) gave a period of 165 days from data obtained in 1955/1956. Our light curve also favours this
period. In between these two measurements are the light curve data presented by Fox (\cite{Fox82}) giving the shorter period. The star therefore
switched its pulsation mode back and forth within about 50 years. This seems to be hardly compatible with the longer timescales
expected from evolutionary models. 
The mode switches seem to be concentrated in the
luminosity range close to the RGB tip. 

\subsection{AGB or RGB stars?}\label{agbrgb}

Throughout this paper we have assumed that all the variables discussed here are on the AGB. As can be seen from
Fig.\,\ref{f:pld} a significant part of our sample is located below the tip of the red giant branch (RGB). 
The RGB tip plotted in Fig.\,\ref{f:pld} is taken from the recent study of Ferraro et al.\,(\cite{FMOF00}).
They give a value of K$_{TRGB}$$=$6.75$\pm$0.2\,mag. This result is in good agreement with
the theoretical value expected for the metallicity of 47\,Tuc (Salaris \& Cassisi \cite{SC98}). All stars are
above the low luminosity limit for thermally pulsing stars, so that they could belong to either the RGB or the
AGB.  The stars well above the RGB tip are clearly AGB stars, namely V1, V2, V3. 
V4 and V8 are both fundamental mode pulsators slightly above the RGB tip and thus they  % moved this sentence here from later in the section.
can be counted as AGB stars, too. 
For the other stars of our sample, we cannot decide on their AGB nature from
luminosity alone. Colour information may be a good indicator for a separation, but our photometry is not
homogeneous enough to decide on this rather small effect as a mean colour over the light cycle would be
required.

Can variability properties allow us to distinguish between AGB and RGB stars?
Using artificial luminosity functions,
Alves et al.\,(\cite{alves98}) and Wood et al.\,(\cite{Wood99}) argued that most
pulsating stars below the RGB tip are in fact AGB stars. However, using
data more sensitive to smaller amplitude variables, 
Ita et al.\,(\cite{ita02}) and Kiss \& Bedding (\cite{kb03}) 
have shown that a substantial fraction of variables below the RGB tip should be RGB stars. 

According to the logP-K-diagram for LMC variables given by Kiss \& Bedding (\cite{kb03}, see their Fig.\,4), 
most of the
variables detected below the tip of the RGB are pulsating in the second or third overtone mode. Furthermore,
Kiss \& Bedding split their logP-K-diagram into six parts according to the light amplitude in $I$. Most
of the variables below the tip of the RGB show amplitudes between 0.01 (their detection limit) and 0.14\,mag.
The luminosity function for these stars indicates a substantial fraction are on the RGB.
However, for larger light amplitudes, the RGB component is no longer detectable. 
Kiss \& Bedding also give colour information showing that
by far the largest fraction of stars below the RGB tip with amplitudes $>$0.14\,mag is found on the blue side
of the red giant branch, which again favours attributing these stars to the AGB.  Thus larger light
amplitudes seem an indication of AGB status.

Kiss \& Bedding (\cite{kb03}) used $I$ band
data, whereas we have V band light curves. Fox (\cite{Fox82}) presents $V$ and $I$
light curves for several of our sample stars. For the stars of lower luminosity like
V5 or V6, the $I$ amplitude is about a factor of 2 less than the $V$ amplitude. 
The $V$ band amplitudes of the stars of our sample are all $\ge$0.2\,mag,
i.e.~we estimate an $I$ amplitude of $\ge$0.1\,mag.

V21, V5 and V6 all show light amplitudes of several tenth of
a magnitude. Such amplitude values favour an AGB nature for these stars as well. V7 shows a
rather small light amplitude, and for V11 the parameters of the light change are not clear.
Therefore, we cannot decide if these two objects are RGB or AGB stars. Interestingly, both
objects also have smaller velocity amplitudes.

Another unclear case is V13. The total amplitude measured by us is $\approx$0.7\,mag. But this amplitude is
dominated by the long period variation, the short period variation has a much smaller amplitude. The amplitudes
used in the discussion by Kiss \& Bedding are derived from Fourier analysis of the individual frequency components.
Therefore the amplitude of the short period is the relevant one in our argumentation. As a result V13 may well
be a RGB star. 

Our approach is based on statistical arguments and not on a complete understanding
of RGB pulsation.  In particular, we note that the LMC stars of Kiss \& Bedding (\cite{kb03})
are likely to be more massive than the 47 Tuc stars and their amplitude behaviour
may be quite different to that of 47 Tuc stars.
We therefore cannot exclude the possibility that some of the variables below the RGB tip
are on the RGB, but for many of them a location on the AGB seems to be more likely. 

\subsection{Comparison with mass loss}\label{massloss2}

Using the velocity amplitude we can separate the logP-K-diagram (Fig.\,\ref{f:pld}) of 47\,Tuc into three to four regions.
At highest luminosity we find the Mira variables (V1, V2 and V3) with the largest velocity amplitudes. 
Going down sequence C, we find the second group (V4 and V8)
with intermediate amplitudes.  The third group consists of stars with low velocity amplitude on sequence B (and possibly A).
Finally, V13 may form a special case with its large amplitude due to a long secondary period.

Mass loss data available in the literature are limited to measurements of circumstellar dust. As there may
be a dust-free mass loss as well, e.g.~during the RGB phase, we cannot rule out that stars with 
indications for no or only small amounts of circumstellar dust still have a considerable mass loss rate.
However, it is very likely that the existence of dust enhances the mass loss rate 
(e.g.\,H\"ofner et al.\,\cite{Hoefner96}),
so we expect that the infrared excess is at least a measure for the {\it relative} mass loss rate within the cluster. 
As a caveat we have to mention the 
lack of infrared photometry beyond 12\,$\mu$m for
the stars of our sample. This certainly limits the reliability of estimates for the amount of dust surrounding a star.

Note that the circumstellar material currently around each star may 
originate from a time when the star's pulsation was different.  Another factor to keep in mind is that it is well known 
(e.g. Olofsson et al.\,\cite{Olof00})
that, at least in some cases, mass loss in AGB stars may be episodic rather than continuous.  This has also been noted in 
the context of
globular clusters by Origlia et al.~(\cite{Origlia02}).

According to Reimers' (\cite{Reimers75}) law, we would expect the mass loss rate to increase with luminosity. Therefore it is
difficult to separate the effects of luminosity and pulsation (velocity amplitude) on the mass loss rate:
high mass loss rates are found at high luminosities, where we also find large velocity amplitudes. We can, however,
try to compare mass loss rate and velocity amplitude for some stars of similar luminosity. Such a group of stars
is found around the RGB tip (V8, V4, V11 and V21). Origlia et al.~(\cite{Origlia02}) measure $K-$[12] values for V8 and V21. V8 shows a
clear dust excess while V21 does not. Comparing the velocity amplitudes for these two stars shows that V8 also has a
significantly larger velocity amplitude than V21. V4 is also a star with an infrared dust excess (Frogel \& Elias 
\cite{FE88}).
It has a similar velocity amplitude as V8. Gillett et al.~(\cite{Gillett88}) report an infrared excess for V11,
but a separation between V11 and the nearby variable V18 is very difficult on the IRAS images they used. Ramdani \& Jorissen
(\cite{RJ01}) later found an infrared excess of V18 from ISO observations, while V11 has only
a very small infrared excess. We conclude that, at similar luminosity, stars with higher pulsation amplitudes show higher mass loss rates.

V5, V6 and V18 form a second group of variables at similar luminosity.  Unlike the group of stars around the RGB tip,
these stars all have very similar velocity amplitudes. Neither V5 nor V6 has an infrared excess reported in the
literature. But V18 has an outstanding infrared excess according to Ramdani \& Jorissen (\cite{RJ01}). 
As noted above, the high infrared excess of V18 at a relatively low luminosity suggests that
this star is currently in the luminosity minimum following a thermal pulse on the AGB.

V13, the star with the long secondary period, shows no indication of 
circumstellar dust. It appears that these long periods on sequence D are not directly related to a mass loss 
phenomenon (compare also Wood et al.\,\cite{Wood04}). 

Origlia et al.\,(\cite{Origlia02}) detected five further stars nearer to the core of 47\,Tuc that show considerable infrared excesses
but that are not known to be variable star. We checked our photometric monitoring data covering two months in late 2003 for variable 
stars at the positions
given by Origlia et al. Four of the five stars had variable counterparts on our V frames. The fifth object is the one with the lowest mass 
loss rate in the list of Origlia et al. 
All four objects varied during the two months by several tenths of a magnitude. More detailed results on these stars including
period determination will be presented in a forthcoming paper (Lebzelter \& Wood in prep.).

Summarizing our results regarding the correlation of mass loss and velocity amplitude: Stars with significant mass loss 
are all pulsating, a fact already noted by other authors (e.g.\,Ramdani \& Jorissen\,\cite{RJ01}). 
Stars on sequence C combine large 
velocity amplitudes with comparably high mass loss rates. On sequence B all stars have low velocity amplitudes and most of 
them also have no or 
very modest mass loss. For stars of similar luminosity, increased pulsational amplitude seems to significantly increase the mass loss rate. 
One star (V18) {\em currently} showing small pulsation
amplitude has considerable amounts of circumstellar dust, presumably from a past interval of higher amplitude pulsation.
We speculated that V18 is 
currently doing a loop in the logP-K-diagram as a consequence of
a recent or ongoing thermal pulse. Its circumstellar dust may then originate not only from a phase of higher
amplitude pulsation but also from a time the star was more luminous.

\begin{acknowledgements}
TL has been supported by the Austrian Academy of Science (APART programme). 
PRW has been partially supported by a grant from the Australian Research Council.
This research at Tennessee State University was partially funded by NASA grant NCC5-511 and NSF grant
HRD-9706268.
This publication makes use of data products from the Two Micron All Sky Survey, which is a joint project of the University of Massachusetts 
and the Infrared Processing and Analysis Center/California Institute of Technology, funded by the National Aeronautics and Space Administration 
and the National Science Foundation.
Partly based on observations
obtained at the Gemini Observatory, which is operated by the Association
of Universities for Research in Astronomy, Inc., under a cooperative
agreement with the NSF on behalf of the Gemini partnership: the National
Science Foundation (United States), the Particle Physics and Astronomy
Research Council (United Kingdom), the National Research Council (Canada),
CONICYT (Chile), the Australian Research Council (Australia),
CNPq (Brazil), and CONICRT (Argentina). Partly based on observations
obtained with the Phoenix infrared spectrograph, developed and operated
by the National Optical Astronomy Observatory. We wish to thank the referee Jacco van Loon
for his helpful comments.

\end{acknowledgements}

\end{document}